\documentclass{article}
\usepackage{graphicx} 
\usepackage{amsmath,amssymb}
\usepackage{upgreek}
\usepackage{hyperref}
\usepackage[letterpaper,margin=1in]{geometry}
\usepackage{authblk}

\title{Characterization of proton-induced damage in thick, p-channel skipper-CCDs}
\author[a]{Brenda A. Cervantes-Vergara\thanks{bcervant@fnal.gov}}
\author[b,c]{Santiago E. Perez}
\author[a]{Claudio Chavez}
\author[d]{Fernando Chierchie}
\author[e]{Brandon Roach}
\author[a,f]{Juan Estrada}
\author[a,e,f]{Alex Drlica-Wagner}

\affil[a]{Fermi National Accelerator Laboratory, Batavia, IL, USA}
\affil[b]{Facultad de Ciencias Exactas y Naturales, Departamento de F{\'i}sica, Universidad de Buenos Aires, Buenos Aires, Argentina}
\affil[c]{Instituto de F{\'i}sica de Buenos Aires (IFIBA), CONICET - Universidad de Buenos Aires, Buenos Aires, Argentina}
\affil[d]{Instituto de Inv. en Ing. El{\'e}ctrica “Alfredo C. Desages” (IEEE), CONICET, Bah{\'i}a Blanca, Argentina}
\affil[e]{Kavli Institute for Cosmological Physics, University of Chicago, Chicago, IL, USA}
\affil[f]{Department of Astronomy and Astrophysics, University of Chicago, Chicago, IL, USA}

\date{}

\begin{document}

\maketitle

\begin{abstract}
In this work, we characterize the radiation-induced damage in two thick, p-channel skipper-CCDs irradiated unbiased and at room temperature with 217-MeV protons. We evaluate the overall performance of the sensors and demonstrate their single-electron/single-photon sensitivity after receiving a fluence on the order of 10$^{10}$~protons/cm$^2$. Using the pocket-pumping technique, we quantify and characterize the proton-induced defects from displacement damage. We report an overall trap density of 0.134~traps/pixel for a displacement damage dose of $2.3\times10^7$~MeV/g. Three main proton-induced trap species were identified, V$_2$, C$_i$O$_i$ and V$_n$O$_m$, and their characteristic trap energies and cross sections were extracted. We found that while divacancies are the most common proton-induced defects, C$_i$O$_i$ defects have a greater impact on charge integrity at typical operating temperatures because their emission-time constants are comparable or larger than typical readout times. To estimate ionization damage, we measure the characteristic output transistor curves. We found no threshold voltage shifts after irradiation. Our results highlight the potential of skipper-CCDs for applications requiring high-radiation tolerance and can be used to find the operating conditions in which effects of radiation-induced damage are mitigated.
\end{abstract}

\section{Introduction}
Charge-Coupled Devices (CCDs), pixelated ionization sensors, are a well-established technology widely used in various scientific applications due to their attractive features, including high quantum efficiency, low dark counts, and low readout noise. Skipper-CCDs combine these features with the added advantage of single-electron/single-photon sensitivity, enabled by their ability to perform multiple non-destructive measurements of the charge in each pixel. This technology has demonstrated excellent performance in applications requiring sensitivity to faint signals, such as direct dark matter detection~\cite{SENSEI2020, sensei2025, DAMICM2023} and astronomical spectroscopy~\cite{Marrufo2024}. To extend its use to scientific applications where detectors are expected to operate in high-radiation environments, such as space missions or accelerator-based experiments, it is essential to demonstrate and quantify its radiation tolerance.

Studies on total dose radiation damage in silicon detectors~\cite{Moll1999, Hartmann2024}, and particularly in CCDs~\cite{Janesick1989, Roy1989, Hopkinson1996, Stefanov2001}, are extensive, and generally classify it into two types: ionization damage and displacement damage. On one hand, ionization damage primarily degrades the insulator layer by creating interface defects and trapped charge within the layer. Interface defects can lead to higher surface dark counts, and trapped charge can lead to voltage shifts causing charge-transfer inefficiency (CTI) between pixels or changes in the operating point of the CCD transistors. On the other hand, displacement damage produces vacancy-interstitial pairs which later form more stable defects within the bulk, such as divacancies and vacancy-impurity complexes, that create intermediate energy levels acting as charge traps. This damage can induce CTI and higher dark counts from hot pixels, the latter primarily promoted by defects with an energy level close to midgap.

The susceptibility of CCDs to ionization damage varies significantly depending on their design and the irradiation conditions. In most cases, the effects of this damage can be mitigated, making displacement damage the primary limiting factor for CCD applications in high-radiation environments~\cite{Hopkinson1996}. Regarding displacement damage, studies indicate that p-channel CCDs are radiation harder than n-channel CCDs due to the characteristics of the induced defects~\cite{Spratt1997, Hopkinson1999, Holland2001, Spratt2005}. The dominant radiation-generated trap in n-channel CCDs is the phosphorous-vacancy (PV) electron trap~\cite{Janesick1991}, while in p-channel CCDs, it is the divacancy (V$_2$) hole trap~\cite{Spratt1997}. P-channel CCDs are considered radiation harder because: 1) the formation of V$_2$ traps in p-channel CCDs is predicted to be less favorable compared to the formation of PV traps in n-channel CCDs~\cite{Spratt2005}, and 2) the PV trap energy ($\sim0.42-0.46$~eV below the conduction band) is closer to midgap than the V$_2$ trap energy ($\sim0.21-0.23$~eV above the valence band), leading to higher dark counts in n-channel CCDs.

\section{Proton irradiation of skipper-CCDs}
\begin{figure}[ht!]
    \centering
    \includegraphics[width=\linewidth]{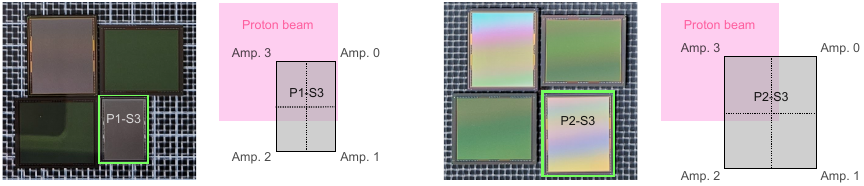}
    \caption{Pictures of the irradiated skipper-CCDs (the sensors tested in this work are highlighted in green) and drawings showing the approximate area irradiated by the proton beam (pink) in each sensor (gray).}
    \label{fig:irradCCDs}
\end{figure}

The sensors tested in this work are thick, p-channel skipper-CCDs, fabricated from high-resistivity n-type silicon, designed at the Lawrence Berkeley National Laboratory (LBNL) in collaboration with the Fermi National Accelerator Laboratory (FNAL) during the R\&D phases of the SENSEI and Oscura~\cite{OscuraSensors2023} dark matter experiments, and fabricated in two different foundries. Each sensor features an array of three-phase 15$\times$15~$\mu$m$^2$ pixels, and four floating-gate readout amplifiers, one in each corner. The sensors were irradiated unbiased and at room temperature at the Northwestern Medicine Proton Center in Warrenville, Illinois, in August 2023, using a 217-MeV proton cyclotron; for details, see Ref.~\cite{Warrenville2024}. To ensure a baseline for comparison, at least a region within the sensor active area and a couple of readout amplifiers of each sensor were kept outside the proton beam line during irradiation as shown in Fig.~\ref{fig:irradCCDs}.

Proton fluences were chosen to reach displacement damage doses (DDD) higher than the estimated for the SNAP satellite orbiting at the second Earth-Sun Lagrange point (L2) for six years, i.e. $6.6\times10^{6}$~MeV/g (Si)~\cite{Dawson2008}. We compute the expected DDD for each fluence assuming a non-ionizing energy loss (NIEL) factor of $1.9\times10^{-3}$~MeV~cm$^2$/g for 217-MeV protons in silicon~\cite{SRNIEL}. The expected total ionizing dose (TID) for each fluence was computed assuming an electronic stopping power of 3.456 (3.605)~MeV~cm$^2$/g for 217-MeV protons in silicon (silicon dioxide)~\cite{SRNIEL}. Table~\ref{tab:CCDdetails} summarizes the main characteristics of the sensors tested in this work, the proton fluences they were irradiated with, and their expected damage doses.
\begin{table}[ht!]
    \centering
    \begin{tabular}{c|c|c|c|c|c}
    Sensor ID & Pixel array size & Thickness & Fluence [p/cm$^2$] & DDD [$10^{8}$~MeV/g] & TID [$10^{11}$~MeV/g]\\
    \hline
    P1-S3 & $1022\times682$ & 675~$\mu$m & $1.2\times10^{10}$ & 0.23 (Si) & 0.41 (Si), 0.43 (SiO$_2$)  \\
    P2-S3 & $1278\times1058$ & 725~$\mu$m & $8.4\times10^{10}$ & 1.60 (Si) & 2.90 (Si), 3.03 (SiO$_2$)\\
    \end{tabular}
    \caption{Geometrical characteristics of the tested sensors, the proton fluences they were irradiated with, and their expected damage doses.}
    \label{tab:CCDdetails}
\end{table}

\section{Characterization of proton-induced damage}
The irradiated sensors were packaged and tested at FNAL. For packaging, the sensor and a Kapton flex cable are glued to a silicon substrate, and the sensor is wire-bonded to the cable. The assembly, consisting of the sensor, flex cable, and substrate, is enclosed within a two-piece copper tray, forming a packaged sensor. For testing, each packaged sensor is mechanically attached to the cold tip of a cryocooler located inside a vacuum chamber. The sensor's flex cable connects to a feedthrough connector on the vacuum side of the chamber, which in turn connects to a Low Threshold Acquisition (LTA) electronics board~\cite{Cancelo2021} on the air side. This board is used for sensor control and readout. During testing, the vacuum chamber is evacuated to a pressure of $\sim1\times10^{-5}$~mbar, and the sensor is cooled to typical operating temperatures, i.e. $\sim$150K. We read out the sensors using all four amplifiers, one per CCD quadrant.

\subsection{Overall performance}
\begin{figure}[ht!]
    \centering
    \includegraphics[width=0.495\linewidth]{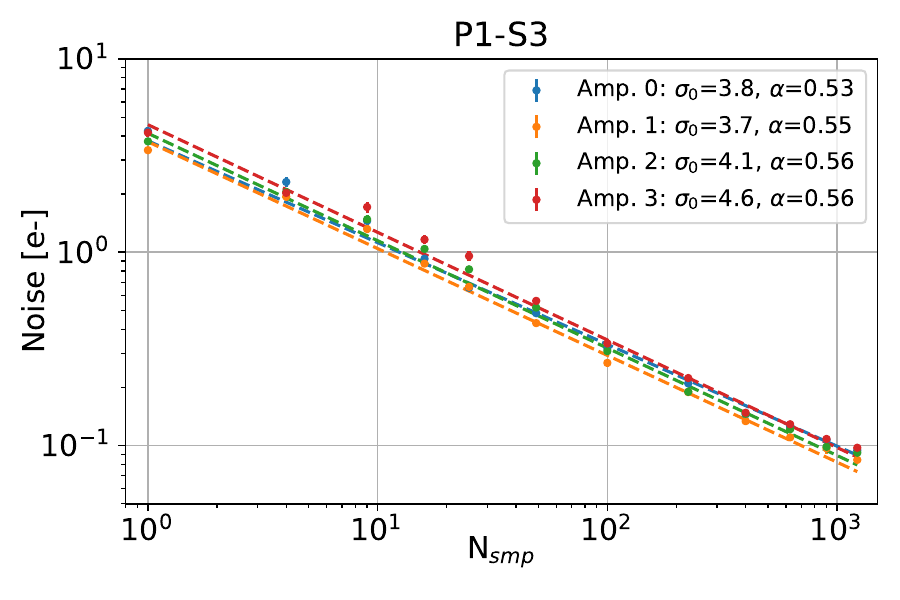}\hfill
    \includegraphics[width=0.495\linewidth]{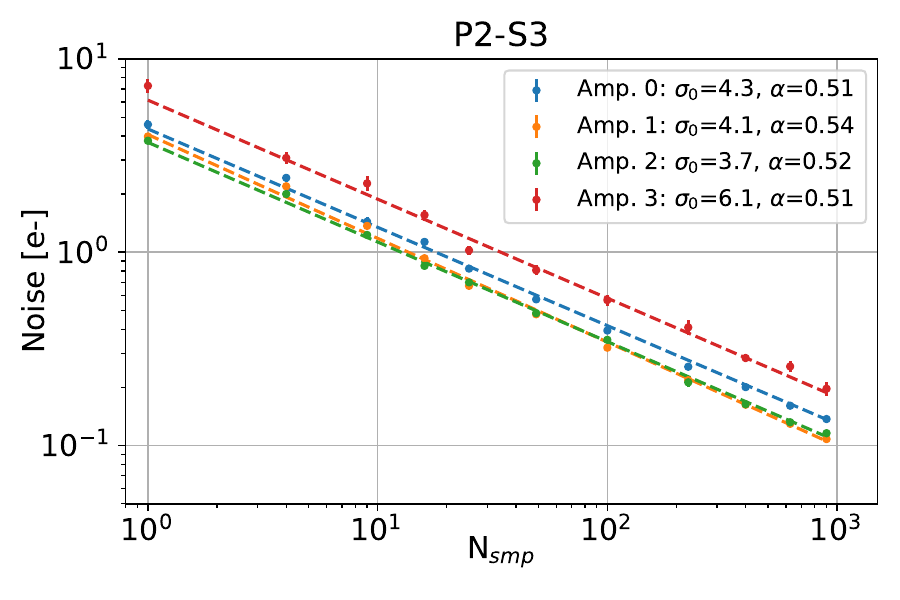}
    \caption{Readout noise as a function of N$_{\rm smp}$ for each amplifier in the irradiated sensors and their corresponding best fit with the function $\sigma_0{(N_{\rm smp})}^{-\alpha}$.}
    \label{fig:noisensamp}
\end{figure}

Dark exposures of the sensors were taken using the typical operating voltages. To evaluate the performance of each amplifier, we tested their ability to perform multiple non-destructive measurements by taking a set of images with different number of samples per pixel $N_{\rm smp}$. The readout noise for each image was computed from the offset-subtracted pixel distribution as the standard deviation of a Gaussian fit centered at zero. We plot this noise as a function of $N_{\rm smp}$ and fit the data with the function $\sigma_0{(N_{\rm smp})}^{-\alpha}$, where $\sigma_0$ and ${\alpha}$ are free parameters representing the single-sample noise and the power-law exponent. The readout noise data and corresponding fits for each amplifier in the tested sensors are shown in Fig.~\ref{fig:noisensamp}.

The readout noise for all amplifiers in both sensors decreased with increasing N$_{\rm smp}$, reaching sub-electron noise levels for N$_{\rm smp}>11$~(35)~samples/pixel in the best (worst) case. In both sensors, amplifier 3 exhibited the highest single-sample noise, comparing the fits. These amplifiers were also located closest to the center of the proton beam area during irradiation. For all amplifiers, $0.51\leq\alpha\leq0.56$, which is expected for low-frequency noise with components $1/\rm f^{n}$ for $1\leq n <2$~\cite{Janesick2001}.

\begin{figure}[ht!]
    \centering
    \includegraphics[width=0.495\linewidth]{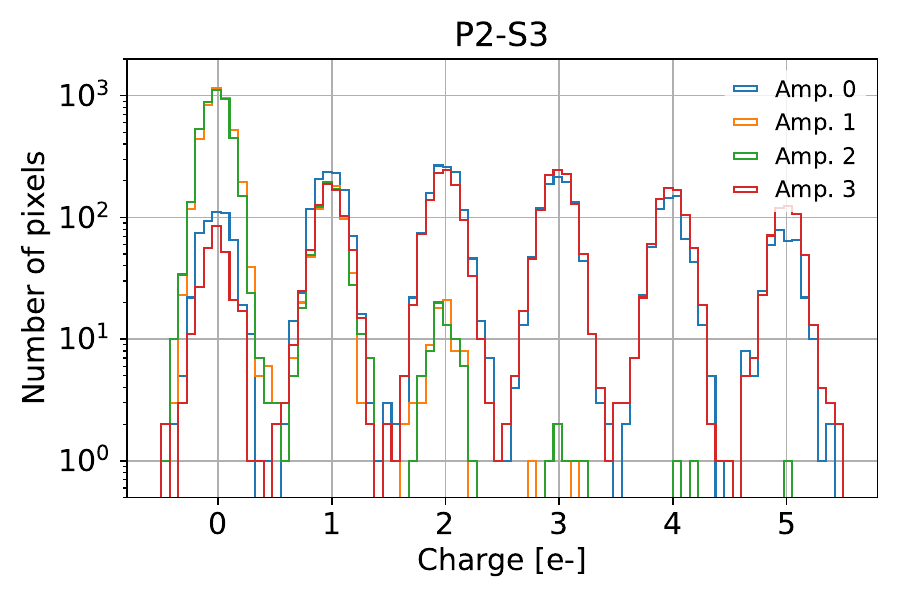}
    \includegraphics[width=0.495\linewidth]{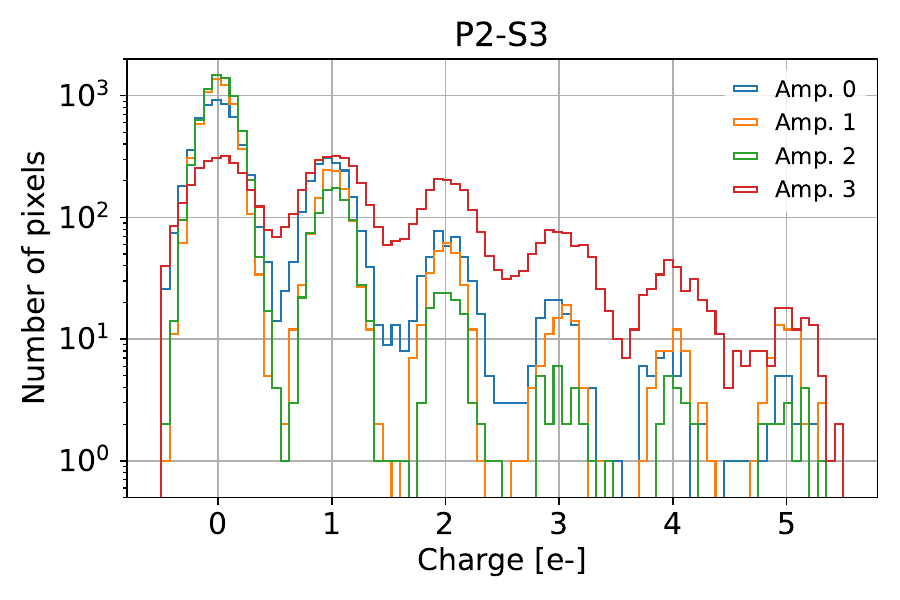}
    \caption{Pixel charge distributions from dark exposure images, acquired with N$_{\rm smp}=625$~samples/pixel, corresponding to data read through each amplifier in sensor P1-S3 (left) and sensor P2-S3 (right). Larger pixel values are evident from regions that were within the beam area during irradiation, i.e. those read through amplifiers 0 and 3 for P1-S3 and through amplifier 0 for P2-S3.}
    \label{fig:pixdists}
\end{figure}

Evident features of radiation damage were clearly observed in the sensors’ active area within the proton beam area when sub-electron resolution was achieved. In standard CCDs, charge trapping and emission of single-electron traps caused by displacement damage manifest as increased dark counts or CTI. In skipper-CCDs, however, the combination of sub-electron energy resolution and the inherent spatial resolution of pixelated sensors reveals individual non-empty pixels with few-electron depositions, enabling precise spatial identification of effects from radiation-induced defects. These effects are also visible in the pixel charge distributions, where discrete peaks at few electrons are more pronounced for the quadrants that were within the beam area during irradiation, i.e. those read through amplifiers 0 and 3 for P1-S3 and through amplifier 0 for P2-S3, as shown in Fig.~\ref{fig:pixdists}.
\begin{figure}[ht!]
    \centering
    \includegraphics[width=0.495\linewidth]{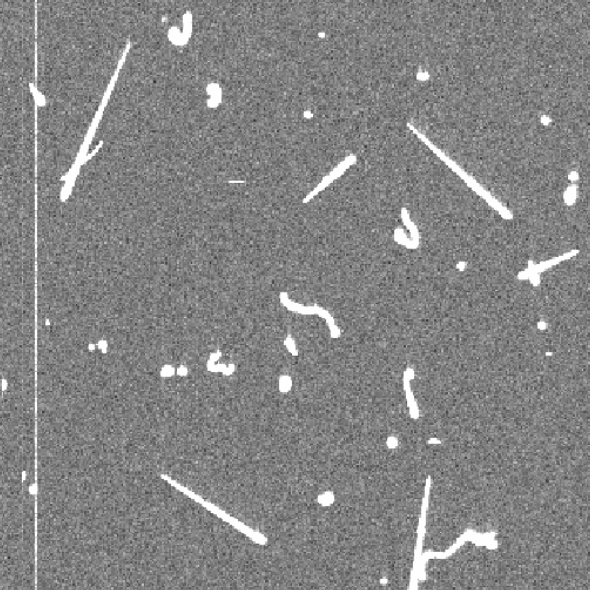}\hfill
    \includegraphics[width=0.495\linewidth]{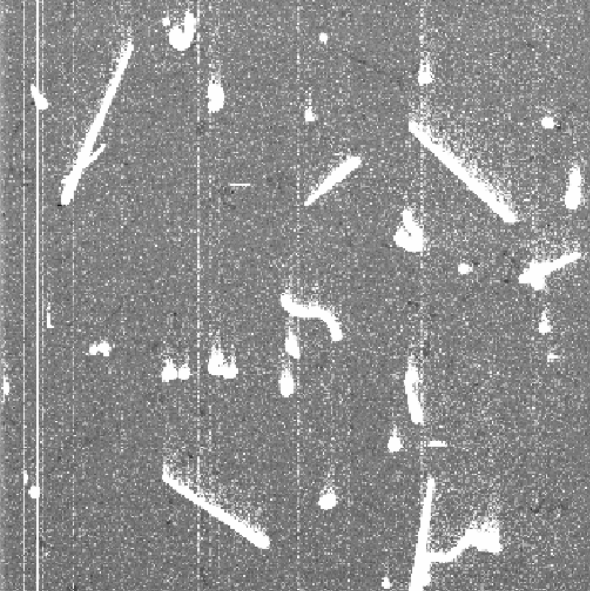}
    \caption{Region of a dark exposure image from sensor P1-S3 (340~pix~$\times$~340~pix) acquired with N$_{\rm smp}=225$~samples/pixel through amplifier 0. In the left image, each pixel value corresponds to a single charge measurement, while in the right image, each pixel value is the average of all 225 measurements. Particle tracks from cosmic radiation interactions are clearly seen in both images, but few-electron depositions from proton-induced damage are only visible in the right image due to the single-electron resolution.}
    \label{fig:stdvsskp}
\end{figure}

\subsection{Displacement damage}
\subsubsection*{Pocket pumping measurements}
Pocket pumping~\cite{Janesick2001, Blouke1988, Hall2014, Bilgi2019} is a powerful technique for spatially localizing and characterizing charge traps in CCDs. It involves repeatedly moving charge between pixel phases, allowing multiple cycles of charge capture and emission that produce identifiable ``dipole'' signals. We use this technique to localize and characterize proton-induced traps from displacement damage, following a procedure analogous to that described in Ref.~\cite{Papertraps2024}. We illuminate the sensor to be tested with a violet LED, providing a relatively uniform charge distribution between 1500–2000~$e^-$/pixel, and we perform an optimized pocket pumping sequence to probe traps under pixel phases 1 and 3. Sensors were read using all four amplifiers with N$_{\rm smp}=10$~samples/pixel. We collected images with $N_{\rm pumps}=2000$ pumping cycles, varying the pumping time $t_{ph}$ between 0.6~$\mu$s to 1~s. We took data at different temperatures, from 140K to 200K.

After subtracting the median pixel value of each row and column in the image, we apply a dipole detection algorithm previously used in Ref.~\cite{Papertraps2024}. This algorithm searches for two consecutive pixels in each column that meet the following criteria: their values have opposite signs and are symmetrical with respect to zero, and their amplitude $\mathcal{A}$, calculated as the absolute value of the difference between their pixel values, is at least five times the median amplitude of any two consecutive pixels in the same column.

\begin{figure}[ht!]
    \centering
    \includegraphics[width=0.495\linewidth]{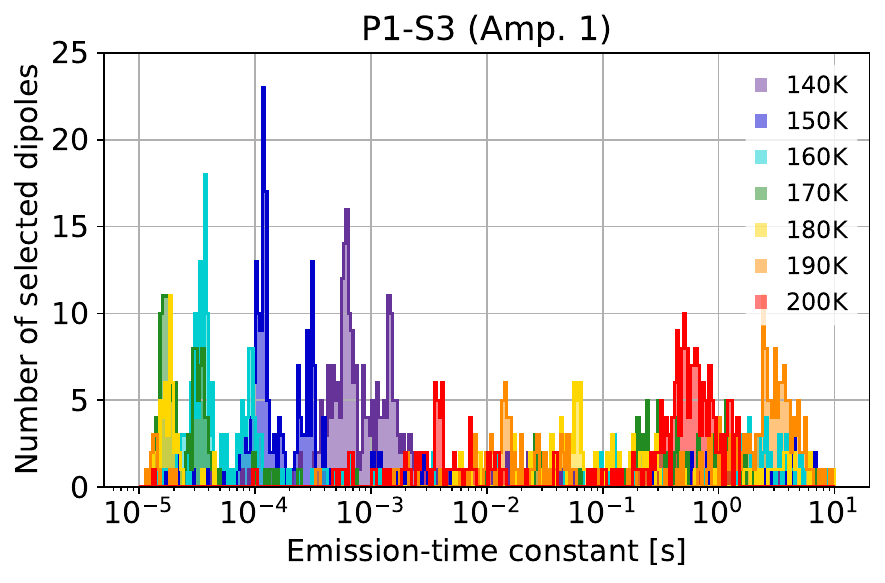}
    \includegraphics[width=0.495\linewidth]{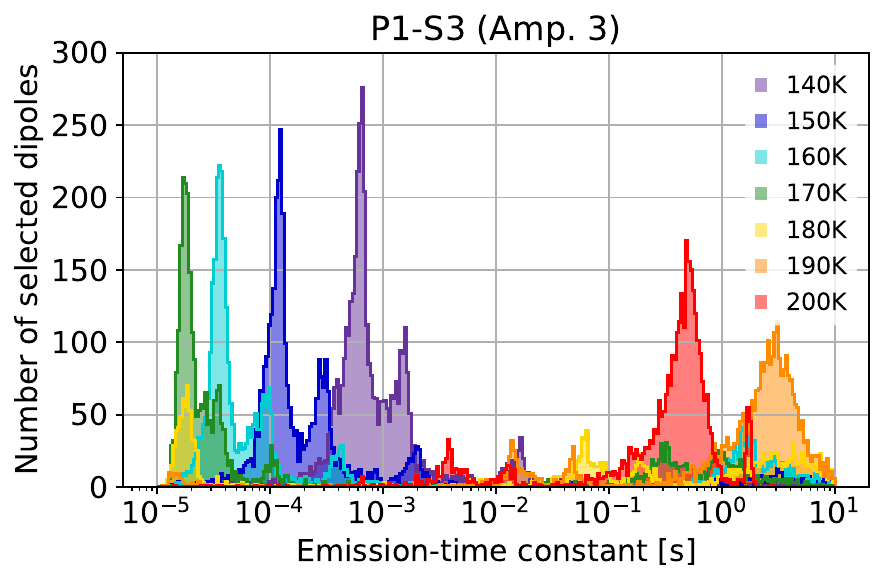}
    \includegraphics[width=0.495\linewidth]{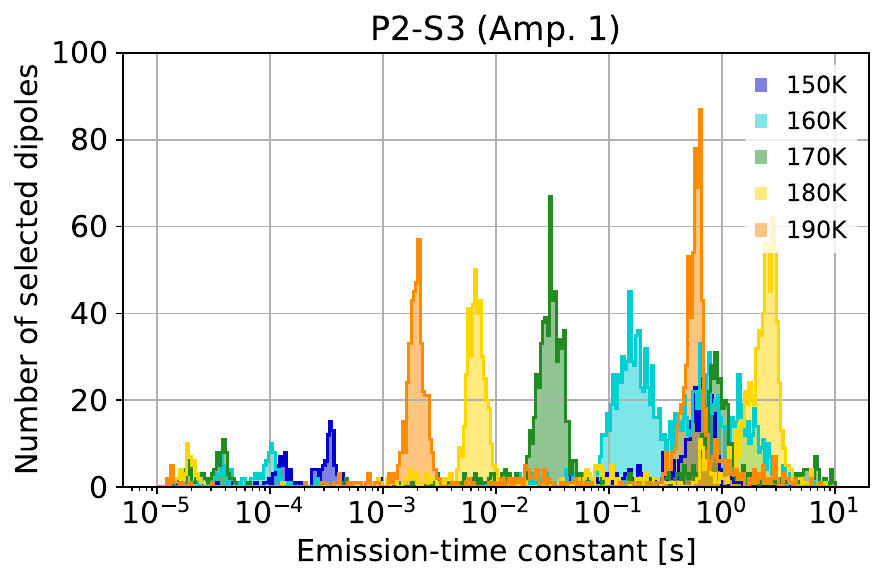}
    \includegraphics[width=0.495\linewidth]{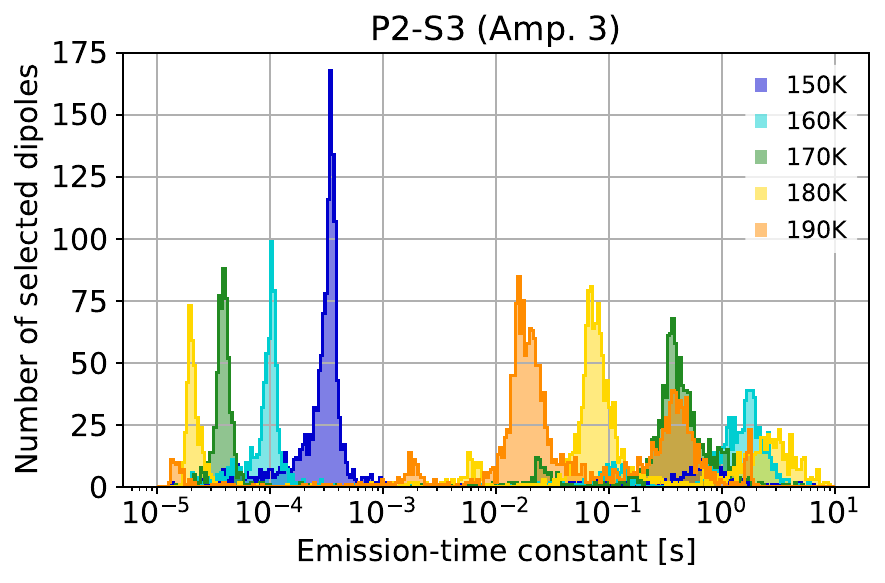}
    \caption{Histograms of the emission-time constants for the detected dipoles that passed the selection cuts in a 240~pix~$\times$~240~pix region next to the readout amplifier at different temperatures. Dipoles were extracted from pocket-pumped images read out through amplifiers 1 (left) and 3 (right) in sensors P1-S3 (top) and P2-S3 (bottom).}
    \label{fig:tauvsT}
\end{figure}

Using sets of images from the same sensor acquired at a fixed temperature, we compute the dipole intensity $I_{\rm dip}=\mathcal{A}/2$ as a function of $t_{ph}$ for each detected dipole. We fit this curve with the following equation,
\begin{equation} \label{eq:dipint}
    I_{\rm dip}=\mathcal{C}\big(e^{-\frac{t_{ph}}{\tau_e}}-e^{-8\frac{t_{ph}}{\tau_e}}\big)\,,
\end{equation}
which is valid for the optimized pocket pumping sequence that we performed~\cite{Bilgi2019}, with $\mathcal{C}$ a constant given by the product of $N_{\rm pumps}$, the trap depth and the probability of charge capture. From the fit, we extract the trap emission-time constant $\tau_e$ associated with the dipole. We select fits with a coefficient of determination greater than 0.7 and a relative error on $\tau_e$ below 50\%. The histograms of the emission-time constants for the detected dipoles that passed the selection cuts, i.e. the ``selected dipoles'', in a 240~pix~$\times$~240~pix region read through amplifiers 1 and 3 in each sensor and at different temperatures are shown in Fig.~\ref{fig:tauvsT}.

In all the $\tau_e$ distributions in Fig.~\ref{fig:tauvsT}, several peaks are observed. The peaks common to both sensors are associated with proton-induced traps, which exhibit emission-time constants ranging from a few microseconds to a few seconds at the tested temperatures. Additional peaks appearing in sensor P2-S3 have been associated in previous work with defects from fabrication~\cite{Papertraps2024}, with emission-time constants from milliseconds to seconds. 

The number of selected dipoles associated with proton radiation damage is higher in the region next to amplifier 3 compared to the region next to amplifier 1 in both sensors. This is expected, as the region next to amplifier 3 was within the proton beam area during irradiation. Although sensor P2-S3 received a higher fluence than sensor P1-S3 during irradiation, resulting in more proton-induced traps, the number of selected dipoles associated with these traps in the histograms in Fig.~\ref{fig:tauvsT} is lower for sensor P2-S3 than for sensor P1-S3. This discrepancy arises from differences in the detection and selection efficiencies of dipoles, which are lower for sensor P2-S3 due to the higher trap density in pocket-pumped images caused by the combined presence of fabrication and proton-induced traps. The difference in efficiencies also explains why the number of selected dipoles corresponding to fabrication traps in sensor P2-S3 is higher for the region next to amplifier 1 than for the region next to amplifier 3, where the significantly higher trap density leads to lower efficiencies.

As the detection efficiency is higher for sensor P1-S3, we estimate the number of proton-induced defects from its data. The highest number of detected dipoles in the whole quadrant read through amplifier 3 was 15,481, obtained from images at 160K. Considering that only traps under pixel phases 1 and 3 were probed, the trap density is estimated to be 0.134~traps/pixel after an irradiation of $1.2\times10^{10}$~protons/cm$^2$ with 217-MeV protons. It should be noted that the pocket pumping technique probes defects under a pixel within an approximate volume of $(10~\mu\textrm{m})^3$ in the buried-channel.

As seen in the histograms in Fig.~\ref{fig:tauvsT}, trap emission-time constants depend on temperature. For p-channel CCDs, according to the Shockley-Read-Hall model for carrier generation and recombination~\cite{Shockley1952}, this dependence is given by
\begin{equation} \label{eq:taue}
    \tau_e=\frac{1}{\sigma v_{th} N_v}e^{\frac{E_t}{k_BT}} \qquad {\rm with} \qquad v_{th}=\sqrt{\frac{3k_BT}{m^h_{\rm cond}}}\qquad {\rm and} \qquad N_v=2\left[2\pi m^h_{\rm dens} \frac{k_BT}{h^2}\right]^{3/2}\,.
\end{equation}
Here, $k_B$ is the Boltzmann constant, $T$ is temperature [K], $E_t$ is the trap energy level [eV], $\sigma$ is the trap cross section [cm$^2$], $v_{th}$ is the charge-carrier's thermal velocity [cm/s], and $N_v$ is the effective density of states in the valence band [cm$^{-3}$]. The dependence on $T$ of $v_{th}$ and $N_v$ is also shown in Eq.~\ref{eq:taue}, where $m^h_{\rm cond}$ is the hole effective mass for conductivity calculations, $m^h_{\rm dens}$ is the hole effective mass for density of states calculations, and $h$ is the Planck constant. Between 100K and 200K, $m^h_{\rm cond}\simeq0.41m_e$ and $m^h_{\rm dens}\simeq0.94m_e$, with $m_e$ the free electron rest mass~\cite{Green1990}.

\begin{figure}[ht!]
    \centering
    \includegraphics[width=0.495\linewidth]{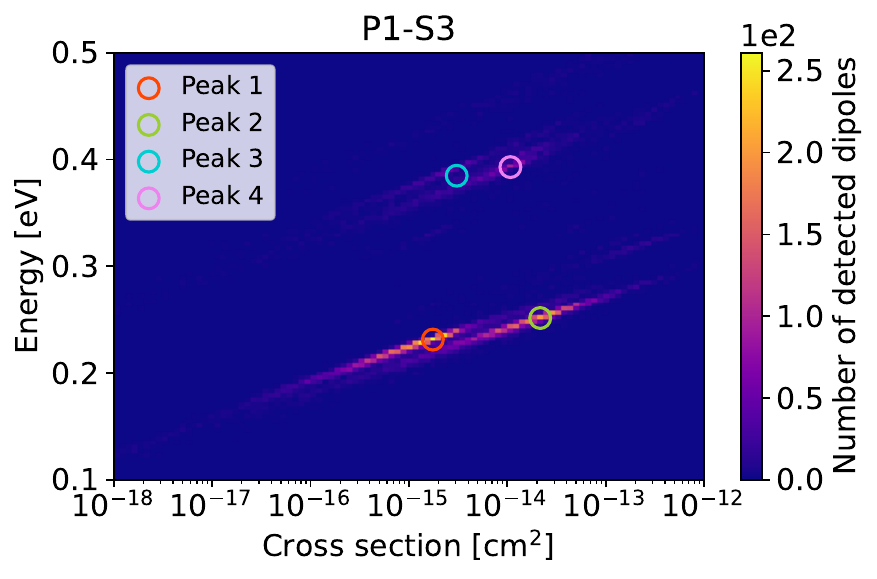} \hfill
    \includegraphics[width=0.495\linewidth]{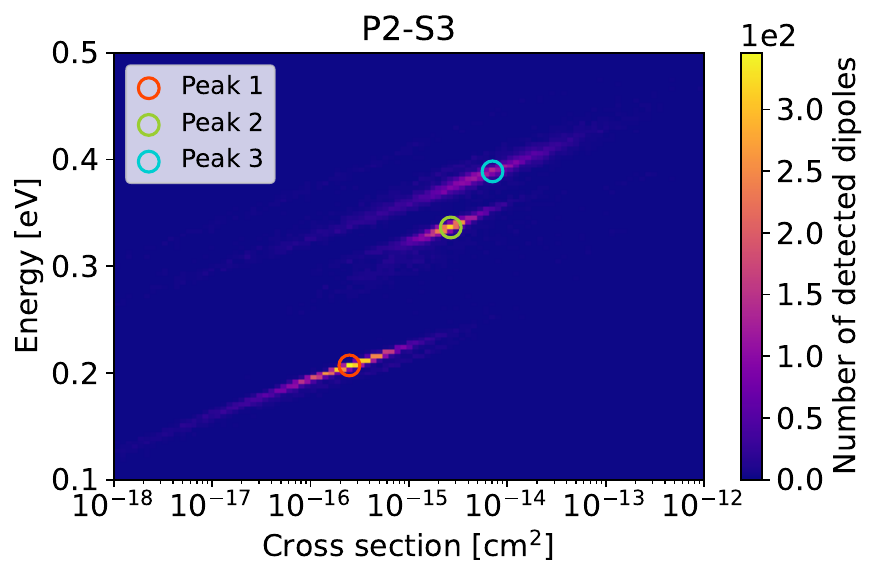}
    \caption{2D histograms of the characteristic trap parameters extracted from the selected fits to $\tau_e$ as a function of $T$ for each selected dipole in sensor P1-S3 (left) and P2-S3 (right). Local maxima associated with different trap species are marked as peaks.}
    \label{fig:2Dhtraps}
\end{figure}

For each selected trap in the histograms in Fig.~\ref{fig:tauvsT}, we plot $\tau_e$ as a function of $T$ and fit it with the function in Eq.~\ref{eq:taue}. From each of those fits, we extract the energy $E_t$ and cross section $\sigma$ associated to each trap. We select fits with a coefficient of determination greater than 0.7 and a relative error on $E_t$ below 50\%. The 2D histograms of the trap energies and cross sections from the selected fits are shown in Fig.~\ref{fig:2Dhtraps}. The local maxima within the histograms are associated with different trap species. From these maxima, we extract the characteristic parameters of the trap species and compare them with values reported in the literature. This information is summarized in Tables~\ref{tab:trapsP1S3} and~\ref{tab:trapsP2S3}.

\begin{table}[ht!]
    \centering
    \begin{tabular}{l|cccc}
    \multicolumn{5}{c}{P1-S3}\\
    \hline
    Peak No. & 1 & 2 & 3 & 4\\
    Energy [eV] ($E_t-E_v$) & 0.231 & 0.252 & 0.389 & 0.393\\
    Cross section [cm$^2$] & $1.8\times10^{-15}$ & $2.2\times10^{-14}$ & $2.7\times10^{-15}$ & $1.1\times10^{-14}$\\
    Trap species identification & V$_2$ & V$_n$O$_m$ (?) & C$_i$O$_i$ & C$_i$O$_i$ (?)\\
    \end{tabular}
    \caption{Characteristic trap parameters associated with each trap species found in sensor P1-S3. Species marked with (?) indicate uncertainty in their identification.}
    \label{tab:trapsP1S3}
\end{table}
\begin{table}[ht!]
    \centering
    \begin{tabular}{l|ccc}
    \multicolumn{4}{c}{P2-S3}\\
    \hline
    Peak No. & 1 & 2 & 3\\
    Energy [eV] ($E_t-E_v$) & 0.207 & 0.336 & 0.389\\
    Cross section [cm$^2$] & $2.5\times10^{-16}$ & $2.7\times10^{-15}$ & $7.1\times10^{-15}$\\
    Trap species identification & V$_2$ & Traps from fabrication & C$_i$O$_i$\\
    \end{tabular}
    \caption{Characteristic trap parameters associated with each trap species found in sensor P2-S3.}
    \label{tab:trapsP2S3}
\end{table}

As discussed in the literature~\cite{Spratt1997, Bebek2002, Mostek2010, Hall2017}, the divacancy is the dominant radiation-generated trap in p-channel CCDs with a donor level reported between $E_v+0.18$ and 0.23~eV and a cross section ranging from $10^{-16}$ and $10^{-15}$~cm$^2$. From our results, we associate this defect with the population of traps related to peak 1 in both sensors, for which we report energies of $E_v+0.21$ and 0.23~eV and cross sections of $2.5\times10^{-16}$ and $1.8\times10^{-15}$~cm$^2$. Also, we find that this trap species has the largest population.

The population of traps related to peak 2 in sensor P1-S3 could be associated with divacancy-oxygen complexes (V$_n$O$_m$). Interstitial oxygen, typically electrically inactive in silicon, is known to effectively trap divacancies, being the divacancy-oxygen interaction the primary mechanism for V$_2$ elimination after annealing at temperatures above 200$^{\circ}$C~\cite{Trauwaert1995, Mikelsen2005}. The V$_2$O hole trap has been reported to have an energy between $E_v+0.23$ and 0.24~eV and a cross section on the order of $10^{-14}$~cm$^2$~\cite{Mostek2010, Trauwaert1995, Mikelsen2005, Markevich2014}. Although our sensors were stored and operated at temperatures between room temperature and $\sim$140K, conditions under which the formation of these defects from annealing is not expected, our measurements yield similar energy ($E_v+0.25$~eV) and cross section ($2.2\times10^{-14}$~cm$^2$) to those reported for V$_2$O.

We identify the traps associated with peak 3 in both sensors as belonging to the same species, with an energy of $E_v+0.39$~eV and cross sections of $2.7\times10^{-15}$~cm$^2$ and $7.1    \times10^{-15}$~cm$^2$. We associate this species to the C$_i$O$_i$ defect, which has been identified in previous work as a radiation-induced defect in p-type silicon~\cite{Bebek2002, Mostek2010}, reporting an energy between $E_v+0.36$ and 0.39~eV and a cross section ranging from $10^{-15}$ and $10^{-14}$~cm$^2$. Traps associated with peak 4 in sensor P1-S3 might also belong to the same species, as our results yield similar energy ($E_v+0.39$~eV) and cross section ($1.1\times10^{-14}$~cm$^2$).

Finally, we identify the traps associated with peak 2 in sensor P2-S3 as traps from fabrication, with an energy of $E_v+0.34$~eV and a cross section of $2.7\times10^{-15}$~cm$^2$. These traps were associated with possible metal impurities in previous work~\cite{Papertraps2024}, reporting an energy of $E_v+0.34$~eV and a cross section of $3\times10^{-15}$~cm$^2$ for sensors from the same fabrication batch as sensor P2-S3.

The slight differences in the trap species energies found in this work compared to reported values in the literature can be attributed to minor variations in the defect structures. The trap species cross sections exhibit higher uncertainty than the energies, which is expected due to their strong dependence on local variables within the lattice, such as the electric field.

\begin{figure}[ht!]
    \centering
    \includegraphics[width=0.495\linewidth]{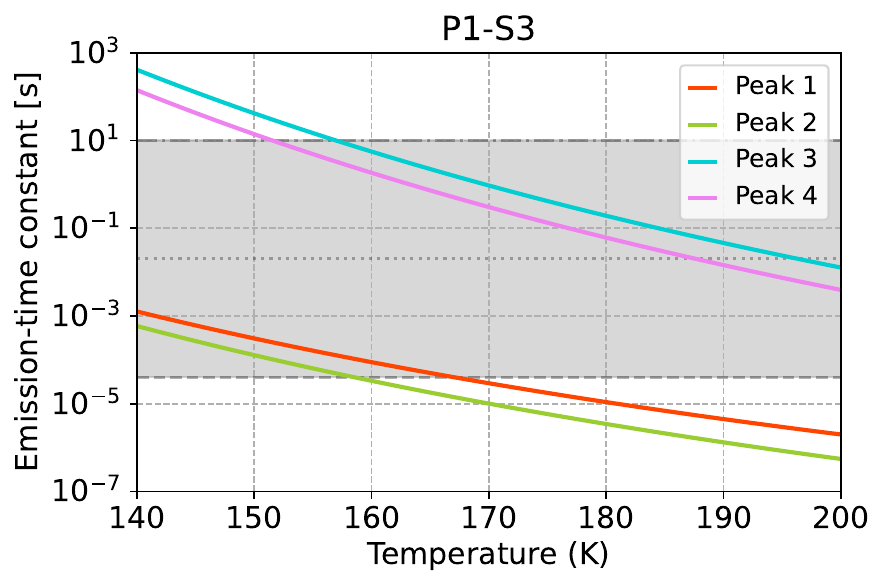} \hfill
    \includegraphics[width=0.495\linewidth]{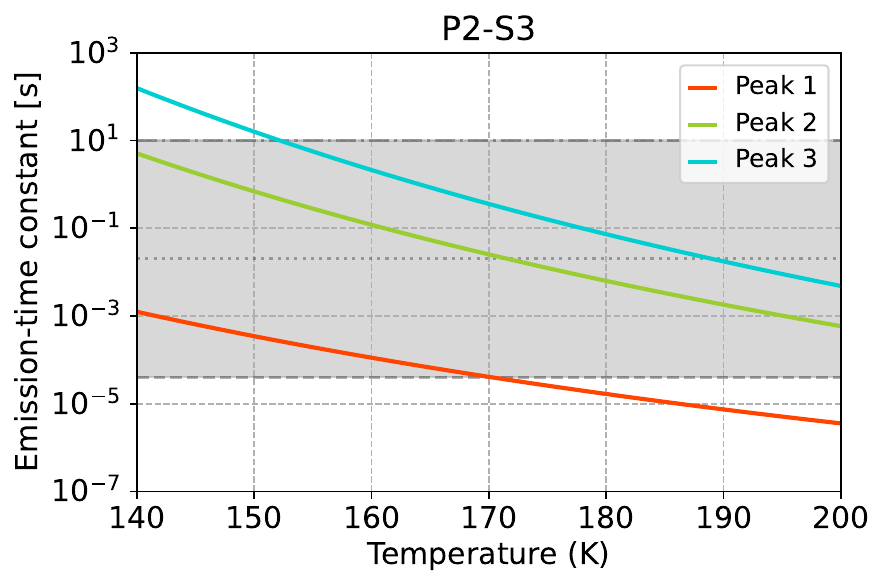}
    \caption{Emission-time constant as a function of $T$ for each of the trap species identified in sensor P1-S3 (left) and P2-S3 (right), using the parameters summarized in Tables~\ref{tab:trapsP1S3} and \ref{tab:trapsP2S3} as input. The gray horizontal lines indicate the time between two consecutive horizontal pixels (dashed), two consecutive vertical pixels (dotted), and the first and last pixels read in an image (dash-dotted), assuming a pixel readout time of 40~$\mu$s and an image size of 500 pix~$\times$~500 pix. Species with emission-time constants within the gray area will emit trapped charge within the image. With larger N$_{\rm smp}$, the gray area shifts linearly toward larger $\tau_e$.}
    \label{fig:tauvsTtraps}
\end{figure}

As discussed in Ref.~\cite{Papertraps2024}, charge depositions from trap emission occur as deferred charge within an image when $t_{\rm pix}<\tau_e<t_{\rm img}$, where $t_{\rm img}$ is the image readout time and $t_{\rm pix}$ is the readout time between two consecutive pixels. Given the dependence of $\tau_e$ on $T$ (Eq.~\ref{eq:taue}), trapped charge is expected to be emitted after more pixels at lower temperatures for a given $t_{\rm pix}$. As also noted in Ref.~\cite{Papertraps2024}, to minimize the distance from the source at which trapped charge is emitted, $\tau_e$ can be decreased by increasing $T$ and/or $t_{\rm pix}$ can be increased by lengthening the CCD clocking times or, in skipper-CCDs, by increasing N$_{\rm smp}$. However, these approaches can lead to increased background from other temperature- and/or exposure-dependent sources, which may be undesirable for certain applications. Fig.~\ref{fig:tauvsTtraps} shows $\tau_e$ as a function of $T$ for each of the trap species identified in each sensor, using the parameters summarized in Tables~\ref{tab:trapsP1S3} and \ref{tab:trapsP2S3} as input. From these plots, we infer that the radiation-induced defect with the greatest impact on charge trapping and detrapping within an image, assuming typical CCD readout times and operating temperatures, is the C$_i$O$_i$. Conversely, although the V$_2$ defect has the largest population, its $\tau_e$ at typical CCD operating temperatures is small enough that is does not produce significant effects from charge trapping and emission processes within an image.

\subsection{Ionization damage}
\subsubsection*{Output transistor curves}
As discussed in the introduction, radiation-induced ionization damage can significantly impact CCD devices by changing the operating parameters of their intrinsic transistors, such as the threshold voltages. To evaluate ionization damage after irradiation, we measure the characteristic curves of the output transistors ($I_{ds}$ as a function of $V_{gs}\equiv V_g-V_s$ and $V_{ds}=V_d-V_s$) in both sensors. Typically, the best operating voltages for the transistors to work as amplifiers are also determined from these curves.

\begin{figure}[ht!]
    \centering
    \includegraphics[width=0.5\linewidth]{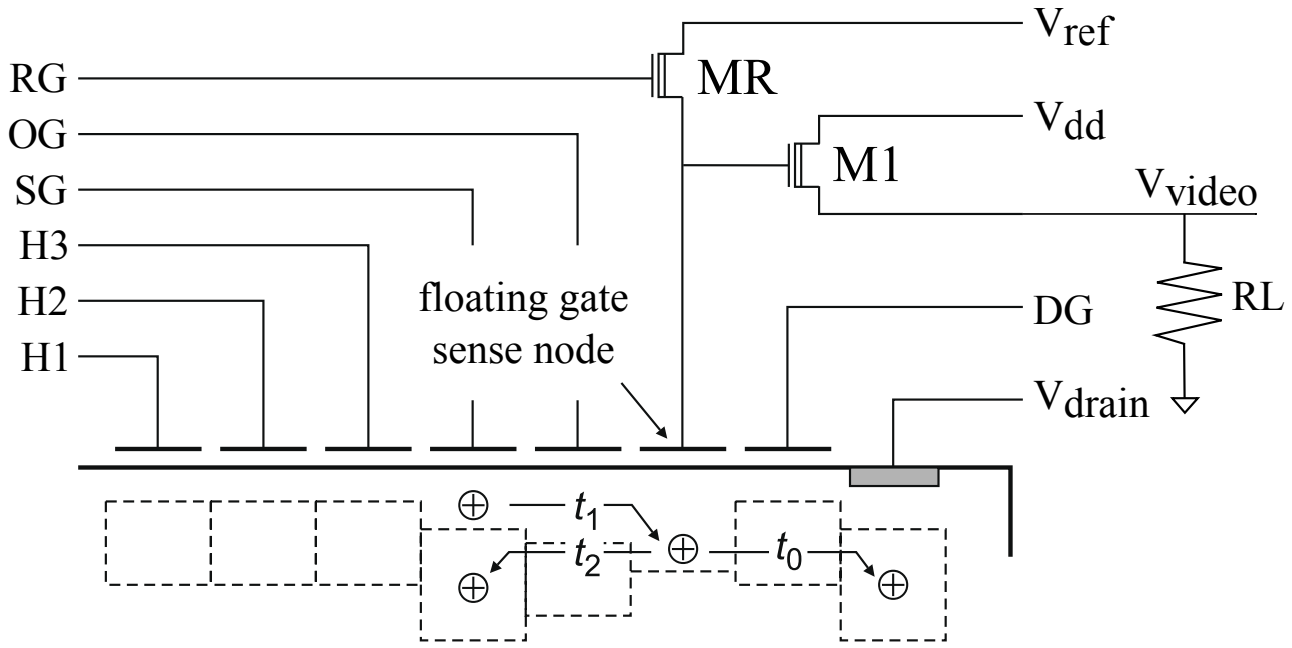}
    \caption{Taken from Ref.~\cite{Tiffenberg2017}. Simplified diagram of the skipper-CCDs output stage.}
    \label{fig:skipperOS}
\end{figure}

A simplified diagram of the output stage of the skipper-CCDs tested in this work is shown in Fig.~\ref{fig:skipperOS}. To measure the characteristic curves of the output transistor (M1), we set the reset transistor (MR) to work in conduction mode so that V$_{\rm ref}\equiv V_g$ in M1. We use an external power supply for V$_{\rm ref}$ to not be limited by the allowed voltage range from the LTA board. For each pair of V$_{\rm dd}\equiv V_d$ and V$_{\rm ref}$, we measure V$_{\rm video}\equiv V_s$. The drain-source current of M1 is computed as $I_{ds}=-V_s$/RL, where RL is a load resistance of 20~k$\Omega$.

\begin{figure}[ht!]
    \centering
    \includegraphics[width=0.495\linewidth]{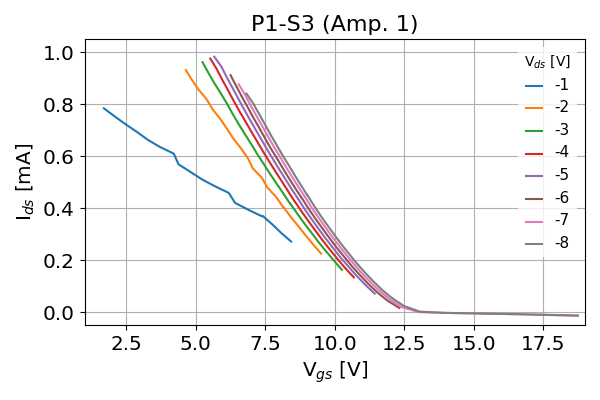} \hfill
    \includegraphics[width=0.495\linewidth]{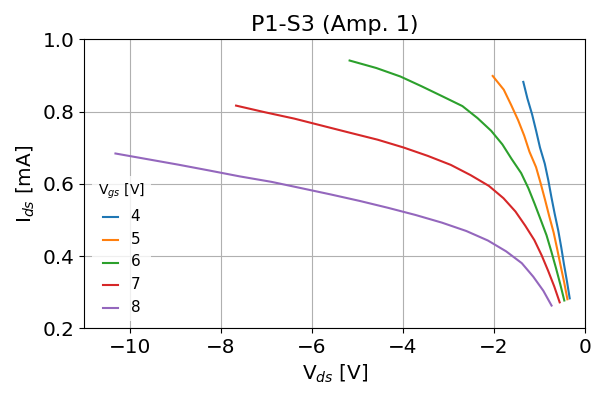}
    \caption{Characteristic curves of the amplifier 1 in sensor P1-S3. Left: $I_{ds}$ as a function of $V_{gs}$ for different $V_{ds}$. Right: $I_{ds}$ as a function of $V_{ds}$ for different $V_{gs}$.}
    \label{fig:curvesP1S3}
\end{figure}

We measure the characteristic curves of all four output transistors in each sensor. During these measurements, the sensors were at 150K and typical operating biases were applied; in particular, the applied substrate voltage was 70V. Fig.~\ref{fig:curvesP1S3} shows the $I_{ds}$ vs. $V_{gs}$ curves for different $V_{ds}$ values (left) and the $I_{ds}$ vs. $V_{ds}$ curves for different $V_{gs}$ values (right) for amplifier 1 in sensor P1-S3, as an example. This amplifier was not within the beam area during irradiation.

\begin{figure}[ht!]
    \centering
    \includegraphics[width=0.495\linewidth]{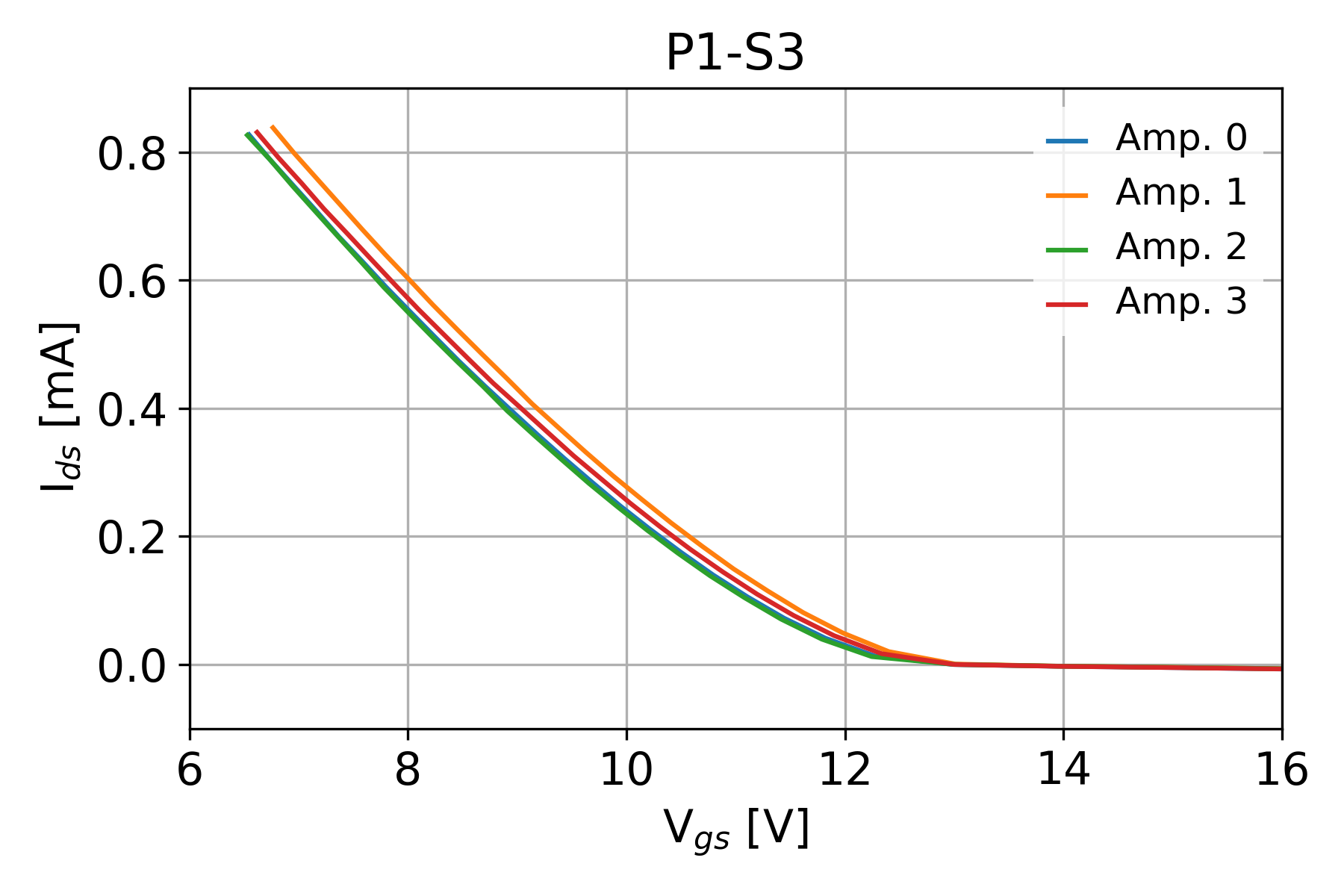} \hfill
    \includegraphics[width=0.495\linewidth]{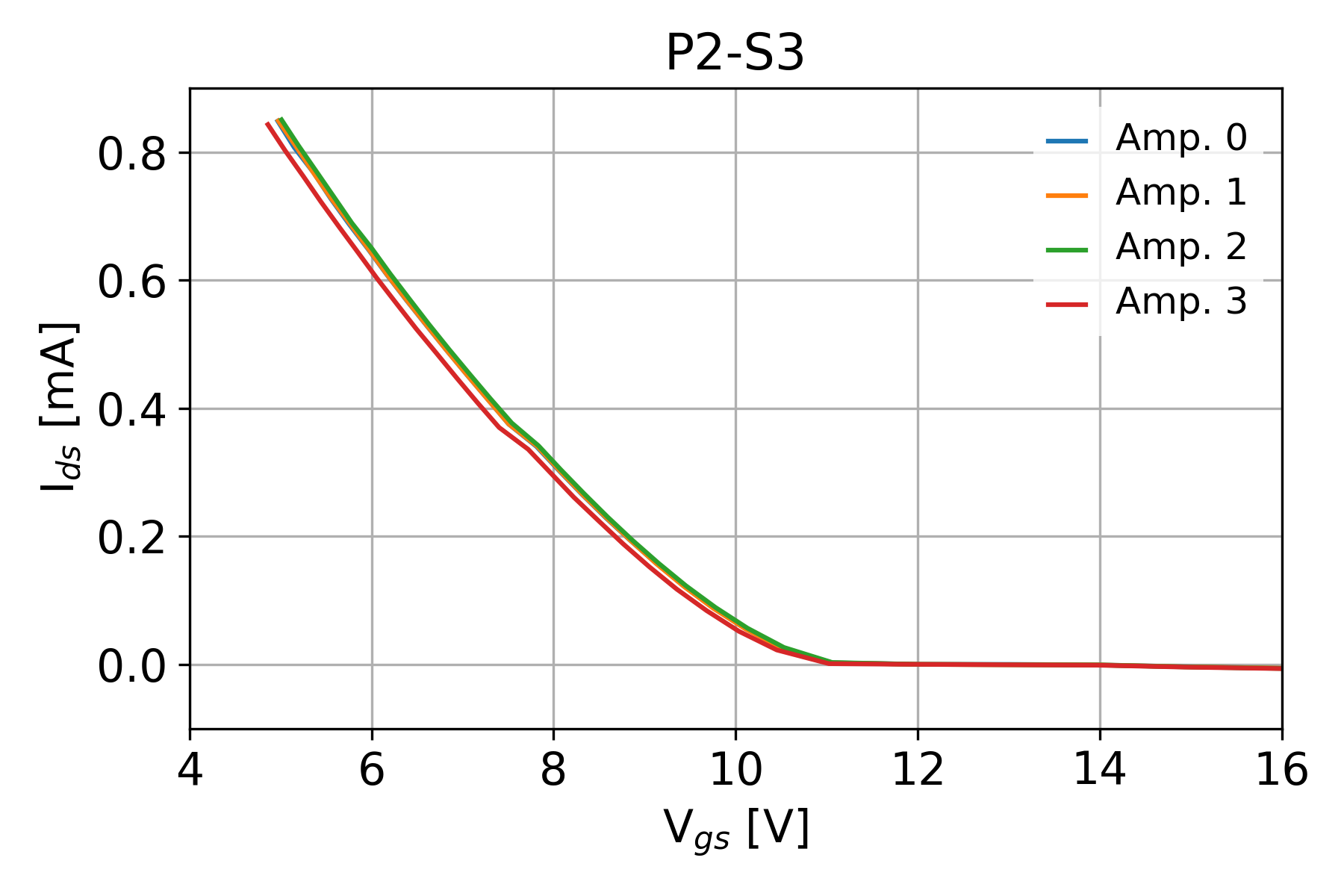}
    \caption{Curves $I_{ds}$ as a function of $V_{gs}$ for $V_{ds}=-7$V of all the amplifiers in sensor P1-S3 (left) and P2-S3 (right). No radiation-induced voltage shifts were observed.}
    \label{fig:transcurves}
\end{figure}

The threshold voltage of the output transistor, $V_t$, can be extracted from the $I_{ds}$ vs. $V_{gs}$ curves, as it is defined as the $V_{gs}$ value for which $I_{ds}=0$. The $I_{ds}$ vs. $V_{gs}$ curves for $V_{ds}=-7$V for all amplifiers in each sensor are shown in Fig.~\ref{fig:transcurves}. We find $V_t$ to be consistent across amplifiers within the same sensor, regardless of whether they were within the beam area during irradiation, with $V_t=(12.8\pm0.3)$V for amplifiers in sensor P1-S3 and $V_t=(11\pm0.2)$V for amplifiers in sensor P2-S3. This consistency can be explained by the nature of voltage shifts, which are associated with trapped charge within the oxide layer. Such trapped charge results from unrecombined holes, and depends on the energy and type of incident radiation, the electron and hole mobilities (which in turn depend on temperature), and the electric field across the oxide. Additionally, trapped charge tends to neutralize at room temperature~\cite{Oldham2003, Schwank2008}. Since the sensors were irradiated unbiased and at room temperature, conditions under which the formation of oxide-trapped charge is unlikely, no significant voltage shifts were observed. The difference in the $V_t$ values between sensors is attributted to differences in the design of the output transistors.

\section{Conclusions}
This work presents the detailed results of the characterization of proton-induced damage on thick, p-channel skipper-CCDs, an important step toward expanding the use of this technology in high-radiation environments. The sensors tested in this work were irradiated unbiased and at room temperature with 217-MeV protons, with fluences on the order of 10$^{10}$~protons/cm$^2$, corresponding to displacement damage doses above 10$^7$~MeV/g.

Despite exposure to high proton fluences, both sensors preserved their ability to perform multiple non-destructive charge measurements, demonstrating single-electron/single-photon sensitivity~\cite{Warrenville2024}. This unique capability of skipper-CCDs, combined with their inherent spatial resolution, enables the precise identification of the effects of radiation-induced defects within an image, which appear as individual pixels with few-electron depositions, features that standard CCDs can only detect as increased charge-transfer inefficiency or higher dark counts.

Using the pocket-pumping technique, we quantified and identified the trap species associated with defects created after proton irradiation. The overall estimated trap density is 0.134~traps/pixel for a displacement damage dose of $2.3\times10^7$~MeV/g. Among the identified trap species, the divacancy, V$_2$, has the largest population in both sensors, with emission-time constants $\tau_e$ below 1~ms for temperatures above 140K. However, its small $\tau_e$ compared to typical readout times minimizes its effects from charge trapping and emission processes within an image. In contrast, the C$_i$O$_i$ defect, also found as trap species in both sensors, exhibited emission-time constants above 4~ms for temperatures below 200K, making it the proton-induced defect with the most significant impact on charge integrity within an image. The trap density associated to this defect should depend strongly on the C and O concentrations within the silicon. Additionally, a third proton-induced trap species, potentially associated with divacancy-oxygen complexes from V$_2$ annealing, was identified in sensor P1-S3. However, this species has emission-time constants at operating temperatures similar to those of the divacancy, resulting in small effects from charge trapping and detrapping. The characteristic parameters associated with each trap species in the irradiated sensors are summarized in Tables~\ref{tab:trapsP1S3} and \ref{tab:trapsP2S3}.

Ionization damage was evaluated through measurements of the output transistor characteristic curves. No significant threshold voltage shifts were found for the output transistors of either sensor. This result is consistent with the conditions under which the sensors were irradiated, i.e. room temperature and unbiased, where oxide-trapped charge formation is unlikely.

The results from this work demonstrate the resilience of the skipper-CCDs output stage to proton irradiation and can be used as a guide to optimize their operating conditions mitigating the effects of radiation-induced damage in future applications, such as the proposed DarkNESS mission~\cite{Darkness2024}.

\section{Acknowledgements}
This work was done using the resources of the Fermi National Accelerator Laboratory (Fermilab), a U.S. Department of Energy, Office of Science, Office of High Energy Physics HEP User Facility. Fermilab is managed by Fermi Forward Discovery Group, LLC, acting under Contract No. 89243024CSC000002. This work was partially supported by NASA APRA award No. 80NSSC22K1411.

\bibliographystyle{ieeetr}
\bibliography{biblio.bib}

\end{document}